\documentclass[reprint,superscriptaddress,aps,prb,twocolumn,floatfix]{revtex4-2}
\usepackage[utf8]{inputenc}
\usepackage{textcomp} % for writing µm
\usepackage{setspace}
\usepackage{amsmath}
\usepackage{breqn}
\usepackage{graphicx}
\usepackage{verbatim}
\usepackage{float}

\usepackage{amsfonts}
\usepackage{amssymb}
\usepackage{xcolor}
\usepackage{soul}
\usepackage{tabularx}
\setcitestyle{super}
\usepackage[colorlinks,linkcolor=blue,anchorcolor=blue,citecolor=blue,urlcolor=black]{hyperref}
\DeclareGraphicsExtensions{.pdf,.eps,.png,.jpg,.mps}
\usepackage{booktabs}
\usepackage{makecell}
\usepackage{threeparttable}
\usepackage{lineno}

\begin{document}

%\pagewiselinenumbers

\title{Integrated adaptive coherent LiDAR for 4D bionic vision}
\author{ Ruixuan Chen$^{1\dagger}$, Yichen Wu$^{1\dagger}$, Ke Zhang$^{2\dagger}$, Chuxin Liu$^{3\dagger}$, Yikun Chen$^{2}$, Wencan Li$^{1}$, Bitao Shen$^{1}$, Zhaoxi Chen$^{2}$, Hanke Feng$^{2}$, Zhangfeng Ge$^{4}$, Yan Zhou$^{4}$, Zihan Tao$^{1}$, Weihan Xu$^{3}$, Yimeng Wang$^{1}$, Pengfei Cai$^{6}$, Dong Pan$^{6}$, Haowen Shu$^{1,5*}$, Linjie Zhou$^{3*}$, Cheng Wang$^{2*}$ and Xingjun Wang$^{1,4,5*}$ \\
\vspace{3pt}
$^1$State Key Laboratory of Advanced Optical Communications System and Networks, School of Electronics, Peking University, 100871 Beijing, China.\\
$^2$Department of Electrical Engineering $\&$ State Key Laboratory of Terahertz and Millimeter Waves, City University of Hong Kong, Hong Kong, China.\\
$^3$State Key Laboratory of Advanced Optical Communication Systems and Networks, Department of Electronic Engineering, Shanghai Jiao Tong University, Shanghai 200240, China.\\
$^4$Peking University Yangtze Delta Institute of Optoelectronics, Nantong 226010, China.\\
$^5$Frontiers Science Center for Nano-optoelectronics, Peking University, 100871 Beijing, China.\\
$^6$SiFotonics Technologies Co., Ltd., Beijing, China.\\
$^\dagger$These authors contributed equally to this work. \\
\vspace{3pt}
\centering{\small Corresponding authors: $^*$haowenshu@pku.edu.cn,$^*$ljzhou@sjtu.edu.cn,$^*$cwang257@cityu.edu.hk, $^*$xjwang@pku.edu.cn}}

%\date{\today}

\maketitle
\noindent
\textbf{Abstract} \\
\textbf{Light detection and ranging (LiDAR) is a ubiquitous tool to provide precise spatial awareness in various perception environments. A bionic LiDAR that can mimic human-like vision by adaptively gazing at selected regions of interest within a broad field of view is crucial to achieve high-resolution imaging in an energy-saving and cost-effective manner. However, current LiDARs based on stacking fixed-wavelength laser arrays and inertial scanning have not been able to achieve the desired dynamic focusing patterns and agile scalability simultaneously. Moreover, the ability to synchronously acquire multi-dimensional physical parameters, including distance, direction, Doppler, and color, through seamless fusion between multiple sensors, still remains elusive in LiDAR. Here, we overcome these limitations and demonstrate a bio-inspired frequency-modulated continuous wave (FMCW) LiDAR system with dynamic and scalable gazing capability. Our chip-scale LiDAR system is built using hybrid integrated photonic solutions, where a frequency-chirped external cavity laser provides broad spectral tunability, while on-chip electro-optic combs with elastic channel spacing allow customizable imaging granularity. Using the dynamic zoom-in capability and the coherent FMCW scheme, we achieve a state-of-the-art resolution of 0.012 degrees, providing up to 15 times the resolution of conventional 3D LiDAR sensors, with 115 equivalent scanning lines and 4D parallel imaging. We further demonstrate cooperative sensing between our adaptive coherent LiDAR and a camera to enable high-resolution color-enhanced machine vision. With its beyond-retinal resolution, flexible gazing ability, as well as low-cost chip manufacturability, our bionic LiDAR solution paves the way for high-resolution adaptive 4D-plus machine vision that could benefit a wide range of applications, including autonomous driving, intelligent robotics, and electric vertical take-off and landing (eVTOL) aircraft.}

\vspace{6pt}
\noindent \textbf{Introduction}\\
\noindent Machine vision is a technology that empowers computers and robots to interpret and analyze visual data from camera images, effectively mimicking human visual capabilities for automation tasks \cite{davies2012computer}. The scope of modern machine vision has expanded beyond traditional cameras to include diverse sensors like LiDAR, with enhanced capabilities through precise distance measurements and 3D environmental understanding \cite{yeong2021sensor}. High-resolution LiDARs facilitate rapid data acquisition and capture fine details in landscapes, which is essential for applications requiring accurate spatial awareness, especially in low-visibility conditions. However, unlike cameras, LiDARs require both laser emission and reception for each pixel, limiting their achievable pixel resolution. The unlimited stacking of physical channels in LiDAR to pursue performance gain is unwise, as it provides a marginal increase in imaging resolution but significantly escalates system complexity and power consumption due to the extra demands of inter-channel synergy. More importantly, simply increasing LiDAR resolution globally is not resource efficient since not all surrounding areas are equally important, and may cause flooded irrelevant information to the system \cite{aeye}.\\ 
%Fig1
\begin{figure*}[ht]
\centering
\includegraphics[width = 18cm]{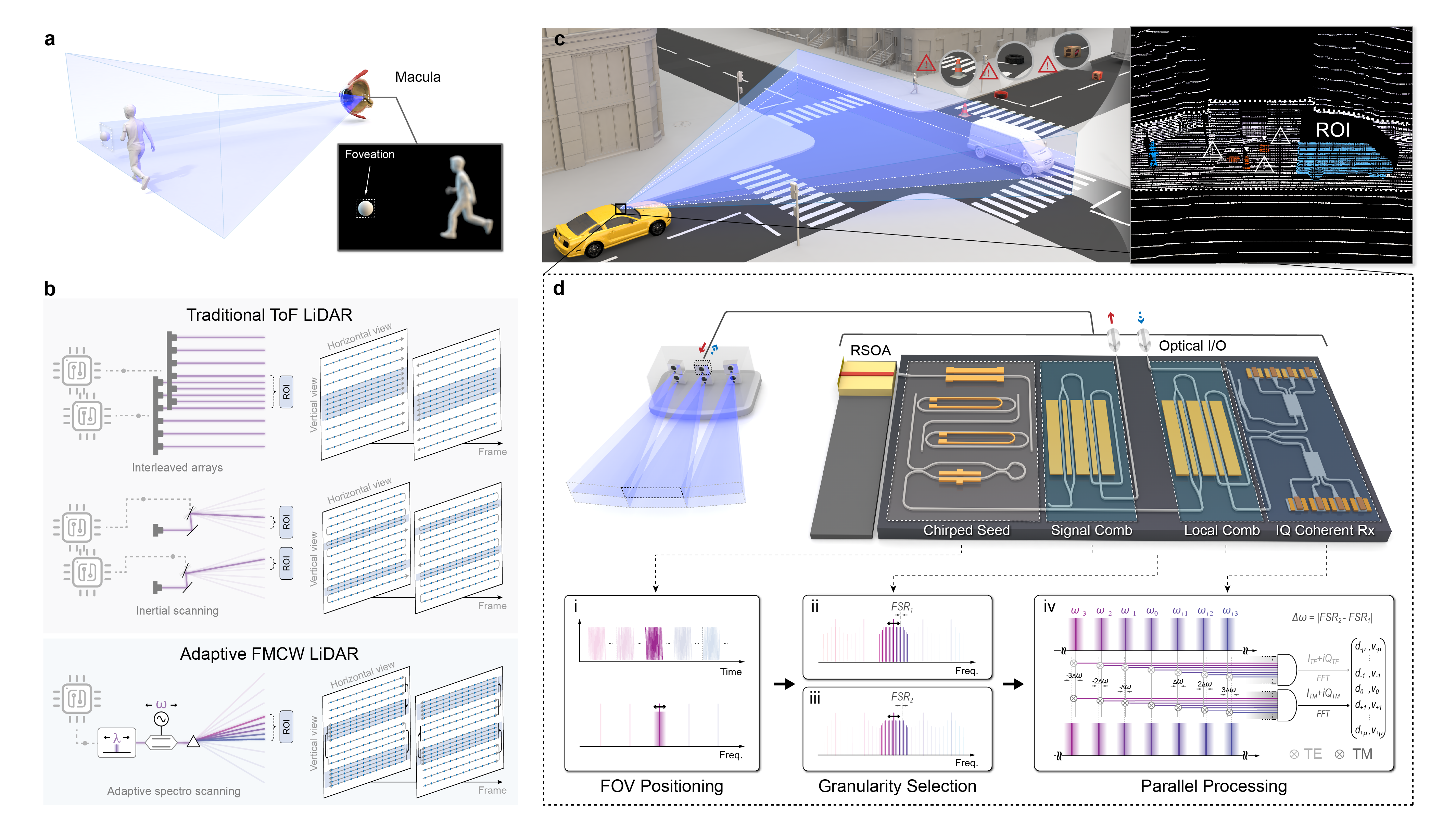}
\caption{\textbf{Adaptive coherent LiDAR for 4D bionic vision. a,} Schematic diagram of retinal foveation vision, where special attention is paid to a certain ROI within a wide field of view. {\bf b,} Schematic comparison between traditional time-of-flight (ToF) LiDAR solutions and the proposed adaptive FMCW LiDAR for the implementation of bionic retinal foveation vision. Bottom: our parallelized solution to realize dynamic gaze. The dark blue regions highlighted in the imaging frames represent ROI. {\bf c,} Conceptual diagram of the adaptive coherent LiDAR applied to urban traffic. Inset: 4D perception with densified point cloud in the ROI to distinguish low-lying obstacles (e.g., traffic cones, tires, cartons). Blue-coded point cloud represents moving objects, and orange-coded point cloud indicates potential road threats. {\bf d,} Schematic illustration of the integrated adaptive coherent LiDAR system. Parallel chirp signals are generated by an electro-optic comb, with the spectral position controlled by the output wavelength of an on-chip external cavity laser (ECL). By tuning the electro-optic comb's repetition rate, scanning patterns with variable granularities can be projected to any desired viewpoints. All wavelength channels are received and processed by an integrated dual-polarized IQ coherent receiver to extract distance and velocity with polarization information. Red arrow: multi-channel probe light; Blue arrow: received echo signal.}
\label{fig1}
\end{figure*}
\indent Nature provides invaluable insights to tackle these challenges, since biological visual systems have evolved to be efficient, energy-saving, and compact. Specifically, human eyes feature a special area in the retina, namely the macula, known for its high concentration of photoreceptors that enable precise and detailed vision \cite{hussey2022patterning}. Coupled with eye movements, greater attention can be paid to any region of interest (ROI) within a broad field of view (FOV) (as shown in Fig.\ref{fig1}a). This provides efficient and high-quality feedback where necessary but without overwhelming the system with data from less relevant areas. Similarly, a “bionic” LiDAR system capable of dynamically focusing on ROIs with higher point densities could enhance perception efficiency in critical areas or search potential threats, thereby reducing hardware resource costs while approaching high-resolution imaging.\\ 
\indent However, blending human-like perception into a compact LiDAR system design presents significant challenges. As shown in Fig.\ref{fig1}b, while traditional multichannel vertical-cavity surface-emitting lasers (VCSELs) can emulate macular vision, they only support static ROIs in fixed interleaved sections between arrays \cite{liang2024evolution}. Currently, dynamic gaze functionalities primarily rely on inertial scanning mechanisms \cite{wang2020mems}, which are susceptible to instability from mechanical fatigue. Although 2-axis MEMS micro-mirrors exhibit low inertia, the trade-off between scanning angle and mirror size restricts the number of channels that can be manipulated. Recent advancements in LiDARs utilizing spectral scanning have shown promise with flexibly programmable imaging patterns, which are based on a temporal-encoded Fourier domain mode-locked (FDML) laser \cite{jiang2020time}. However, the proposed time-stretch scheme operates in serial mode to prevent overlap during wavelength detection with a single-pixel detector, resulting in increased latency since light pulses in each channel must complete its cycle before the next begins. Although spectro-temporal encoding method allows for parallel detection \cite{zang2022ultrafast}, its fixed time-delay configuration \cite{yegnanarayanan1996recirculating} leads to a pre-set and immovable ROI. It also features limited scalability due to the fiber optic systems used. In short, existing LiDAR schemes have not been able to fully replicate the adaptive gaze functionality in human eyes, as such a system should feature scalable parallel detection, adaptive imaging resolution, and adjustable ROI positioning at the same time. Additionally, 4D perception and beyond, with the ability to simultaneously acquire multi-dimensional physical parameters—such as distance, direction, Doppler, and color—through cooperative sensing between multiple sensors, is crucial for achieving comprehensive situational awareness and robust environmental interpretation.
%Fig2
\begin{figure*}[ht]
\centering
\includegraphics[width = 18cm]{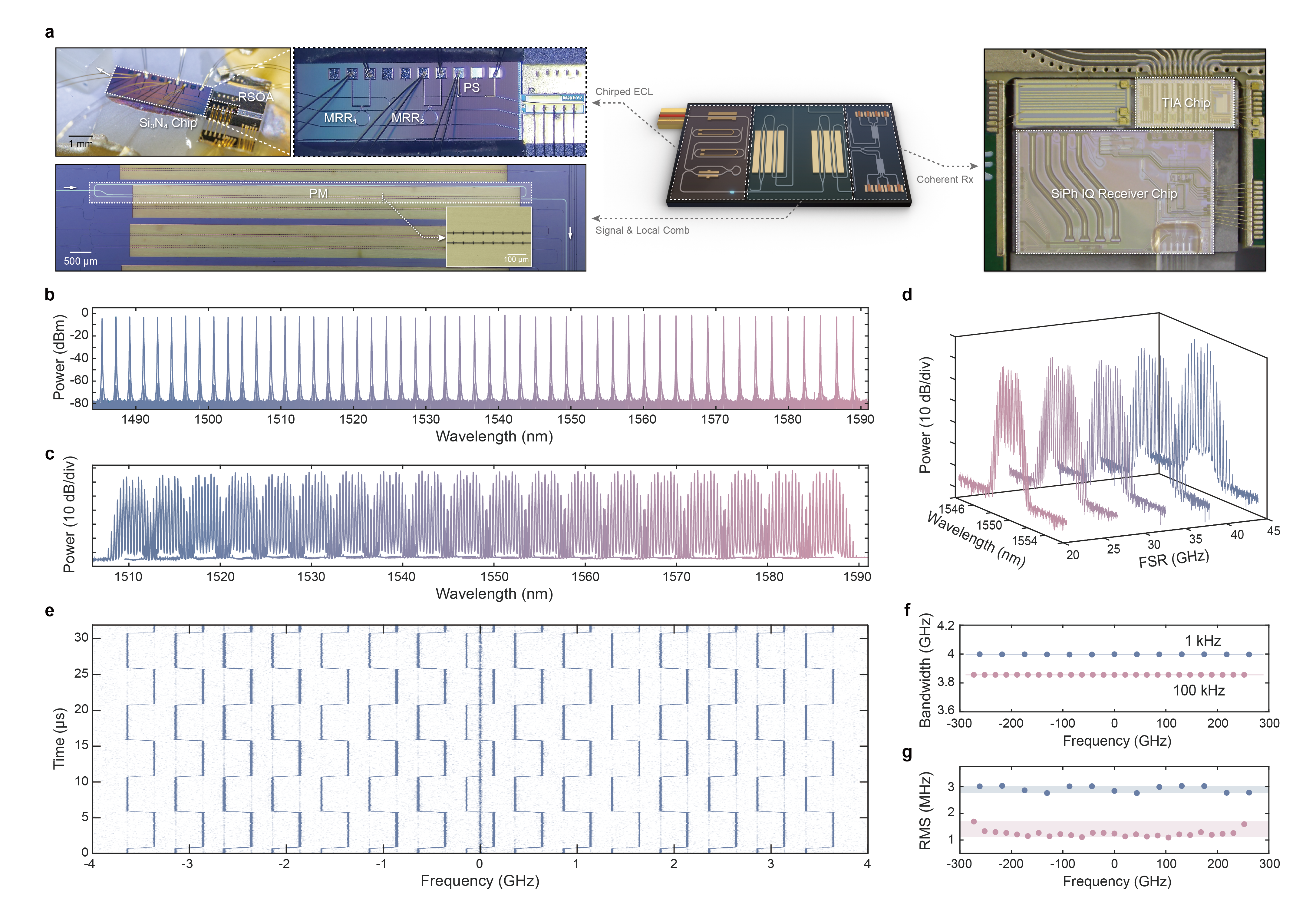}
\caption{\textbf{Adaptive multichannel FMCW signal generation and reception. a,} Photographs of the hybrid integrated ECL (top left), electro-optic comb generator (bottom left) and the silicon photonic (SiPh) IQ receiver (right). The inset shows a zoom-in view of the slotted electrodes of the comb generator. PS: thermally tuned phase shifter; PM: phase modulator. {\bf b,} Superimposed lasing spectra of the hybrid ECL, with the center wavelength of the Vernier filter scanned throughout the tuning range (1486-1590 nm) at 20 ° C. {\bf c,} Broadband electro-optic combs generated at different input wavelengths from 1510 to 1587 nm. {\bf d,} Optical spectra of electro-optic combs driven by optimized RF frequencies (20.4 to 43.9 GHz) at 1550 nm. {\bf e,} Time-frequency spectrogram of multichannel heterodyne beatnotes between a signal comb (21 GHz) and local oscillators (20.5 GHz), tested in a 37 m delayed self-heterodyne fiber optic link. {\bf f,} Channel-dependent chirp bandwidth and {\bf g,} chirp linearity from a perfect triangular frequency chirp for two electro-optic combs (21 GHz and 43.6 GHz) generated at chirp repetition frequency of 1 kHz (blue) and 100 kHz (pink).}
\label{fig2}
\end{figure*}

\noindent\textbf{Concept of integrated adaptive coherent LiDAR for 4D bionic perception} \\
\noindent Here, we demonstrate an integrated coherent LiDAR system that achieves parallel and adaptive imaging simultaneously, leading to a more nuanced and informed machine perception system that can adapt to complex and dynamic environments with low system latency. For example, in a typical urban setting (Fig.\ref{fig1}c), the 4D "eyeonic" perception facilitated by our adaptive coherent LiDAR can quickly scan the  ROIs with high resolution and identify low-lying obstacles, while seamlessly performing semantic segmentation on both static and dynamic point clouds. This resource-efficient methodology is essential for maintaining high frame rates necessary for rapid navigation decision making in such scenarios. The parallelism of our adaptive coherent LiDAR system is achieved by an agile external cavity laser (ECL) pumped electro-optic comb, which generates frequency-modulated continuous laser beams across multiple wavelength channels that can be dispersed toward a broad range of viewing angles (Fig.\ref{fig1}b bottom). This configuration offers broad wavelength tunability and adjustable channel spacing, enabling dynamic reconfiguration of the ROI by adjusting the ECL's output wavelength and the repetition frequency of the electro-optic comb. This flexibility ensures optimal adaptability in diverse environments and applications. As illustrated in Fig.\ref{fig1}d, our hybrid integrated chip-scale system consists of a reconfigurable parallel transmitter and an adaptive parallel receiver. The transmitter is comprised of an on-chip tunable ECL \cite{liu2024highly,lukashchuk2024photonic,wanghigh,lihachev2023frequency,franken2024high} and a signal electro-optic comb generator\cite{zhang2023power,zhang2019broadband,hu2022high,rueda2019resonant}. Both renowned for their versatile tuning capabilities. Leveraging the extensive spectral tuning range of the ECL, the viewpoints can be precisely positioned at any desired location within a broad FOV. Near the selected center wavelengths, a linear triangular frequency-modulated signal from the ECL is distributed to multiple equally spaced frequency channels using an on-chip electro-optic comb, allowing for the simultaneous illumination and detection of multiple pixels. Importantly, the angular imaging resolutions at certain viewpoints can be flexibly controlled by adjusting the electro-optic comb’s repetition rate, which depends on the RF driving frequency. Such adaptability enables the system to focus on any desired viewpoint within the FOV, much like human eye's ability to gaze and concentrate on specific points. \\
\indent On the receiver side, another electro-optic comb generator produces a local oscillator comb from the same ECL source, with a slight repetition rate offset ($\Delta \omega $ in Fig.\ref{fig1}d) from the signal comb, to beat with the received echo signal at a dual-polarized in-phase (I) and quadrature (Q) coherent receiver. This configuration allows multichannel echoes, superimposed with both static and motion data, to be individually resolved and processed at different intermediate frequencies through multi-heterodyne detection  \cite{lukashchuk2022dual,wang2022dual,burghoff2016computational,chomet2024heterodyne} at the coherent receiver. Since both two comb generators are driven by the same ECL output, the down-conversion of signals from optical to radio frequencies simplifies analysis and compensates for wavelength variation arising from adjustable channel spacing or spectrum scanning. Moreover, extracting signals from two orthogonally polarized states enhances perception by providing richer data for texture classification and robust detection. This dual-polarization approach improves resistance to any signal-to-noise ratio (SNR) degradation caused by polarization changes in the return signal, ensuring consistent performance even under complicated conditions.

%Fig3
\begin{figure*}[ht]
\centering
\includegraphics[width = 18cm]{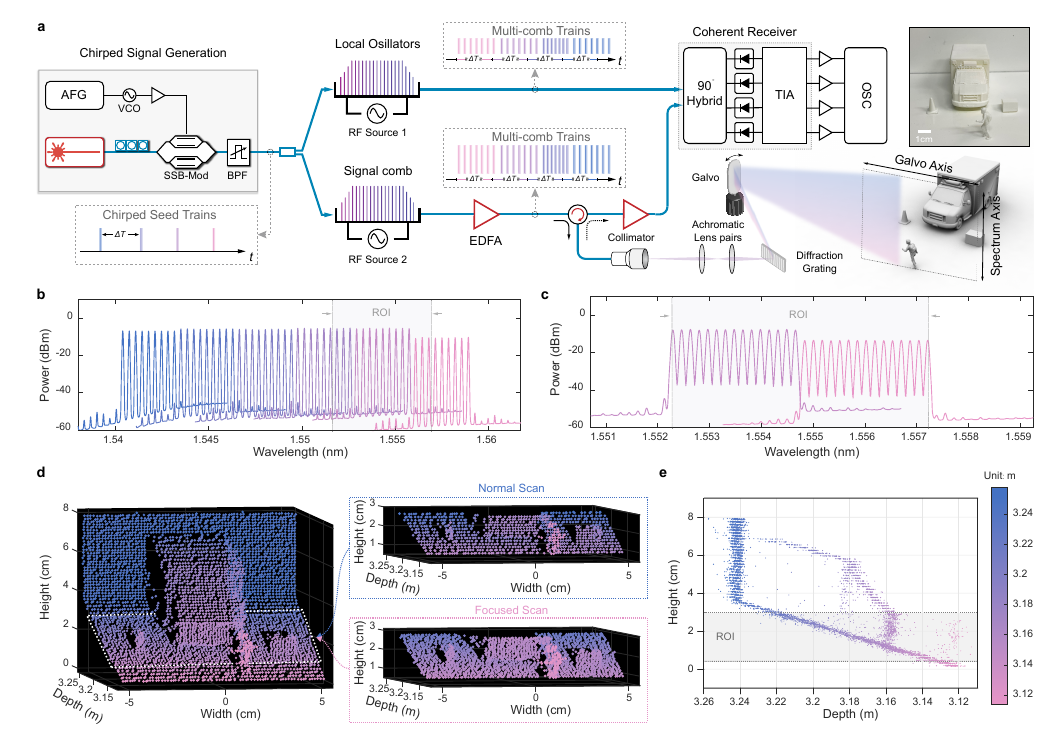}
\caption{\textbf{Ultra-high-resolution gaze imaging. a,} Experiment setup for parallel gaze imaging using a pair of chirped electro-optic combs. The frequency modulation of the chirp seed is achieved through a single sideband modulator (SSB) driven by a voltage-controlled oscillator (VCO). Predistortion of the triangular ramp is performed with an arbitrary function generator (AFG) after linearity optimization. The valid sideband is selected by a bandpass filter (BPF). The inset photograph shows a simulated road scenario with four targets against a whiteboard background. TIA, trans-impedance amplifier; OSC, oscilloscope. \textbf{b,} Optical spectra of six signal combs, with comb lines spaced at 43.5 GHz and driven by a 27 dBm RF amplifier. \textbf{c,} Optical spectra of two densified electro-optic combs for gaze imaging, covering the same ROI highlighted in light gray in \textbf{b}. \textbf{d,} Reconstructed global imaging frame of the simulated road scene, with insets on the right panel showing a partially truncated global image and a gaze image centered on the small obstacle region. \textbf{e,} Depth imaging results in a single horizontal slice of the full LiDAR imaging result shown in \textbf{d}, showing significantly higher imaging resolution in ROI.
}
\label{fig3}
\end{figure*}

\vspace{6pt}
\noindent
\textbf{Adaptive parallel FMCW LiDAR source} \\
\noindent In this work, we achieve adaptive multichannel FMCW signal generation by cascading a hybrid integrated ECL and an electro-optic comb generator. According to the principle of spectral scanning, the imaging FOV and granularity are both controlled by wavelength. To achieve a wide FOV coverage, the hybrid integrated ECL is designed to include an indium-phosphide (InP) reflective semiconductor optical amplifier (RSOA) with a broad gain bandwidth, coupled with a low-loss silicon nitride ($Si_{3}N_{4}$) external cavity chip where the filtered feedback wavelength is widely tunable via a pair of Vernier microring resonators (MRRs), as shown in Fig.\ref{fig2}a. Figure \ref{fig2}b illustrates the superimposed spectra of laser seeds generated at approximately 2 nm intervals, with a wavelength tunability exceeding 100 nm and an overall side mode suppression ratio (SMSR) greater than 70 dB, which serves as a chirped source for the subsequent generation of electro-optic combs at any desired wavelength.\\
\indent The electro-optic comb is generated on a thin-film lithium niobate (TFLN) platform using a dual-pass phase modulator \cite{yu2022integrated}, which enhances modulation efficiency by doubling the interaction length while keeping a compact footprint of only 11.5 mm × 0.6 mm. Slot electrode design \cite{kharel2021breaking,xu20246} is applied for better velocity matching, leading to a high electro-optic bandwidth. The excellent wavelength-reconfigurability of electro-optic combs allows us to realize efficient comb generation at a wide range of input laser wavelengths from 1510 nm to 1587 nm (as shown in Fig.\ref{fig2}c). The electro-optic comb is driven by a 27 dBm RF amplifier at 43.8 GHz repetition rate. The collective spectra reveal that all generated combs across the varying wavelengths maintain uniform characteristics in terms of flatness and the number of generated lines, which is vital for ensuring consistent detection performance in spectral scanning. Importantly, point cloud scanning with different granularities (e.g. finer scan in ROIs) can be achieved by adjusting the repetition rate of the electro-optic comb without changing FOV. In our dual-pass configuration \cite{zhang2023power}, the phase modulation efficiency is maximized at periodically appearing optimal RF frequencies, when the recycled optical signal remains in phase with the driving microwave. Figure \ref{fig2}d illustrates the spectra of the electro-optic combs driven at five optimal RF frequencies from 20.4 GHz to 43.9 GHz, which are identified through electro-optic response measurements, detailed in Supplementary Note \textcolor{magenta}{II}. Although the electro-optic comb envelope generated through pure phase modulation follows the Bessel function of its first kind and is not flat, we can achieve a flat-top electro-optic comb by cascading amplitude and phase modulation (see Supplementary Note \textcolor{magenta}{I}), which would effectively eliminate the need for channel power shaping.\\
\indent For each laser seed, linearized triangular frequency modulation is achieved by applying a pre-distorted sweep waveform to the thermal-optic phase shifters integrated into the ECL’s two MRRs. This efficient modulation process enables broadband chirped signals (2.6 GHz to 4.1 GHz), modulated at a 1 kHz chirp repetition frequency (see Supplementary Note \textcolor{magenta}{III}). Furthermore, the frequency noise measurement results (see Supplementary Note \textcolor{magenta}{IV}) exhibit that the intrinsic linewidths are distributed from 1 kHz to 4 kHz, while the effective $1/\pi$-integral linewidths are around 55 kHz, which is essential for precise and long-range applications. Even faster chirp signal can be generated on the TFLN photonic platform \cite{wanghigh,li2022integrated}, which is well-suited for developing high-performance frequency chirped laser.\\
\indent To demonstrate intrinsic mutual coherence between all comb lines of our parallel FMCW LiDAR source, we used 1 kHz chirped ECL to simultaneously excite two electro-optic combs (spaced at 21 GHz and 20.5 GHz) as signal comb and local oscillators. After traversing optical fibers with a length difference of 37 m, the signals were input into the SiPh coherent receiver for multi-heterodyne detection. Figure \ref{fig2}e shows the time-frequency map of delayed self-heterodyne beat spectroscopy. Since the frequency chirp of ECL is applied to all comb lines, they exhibit identical beat patterns. The chirp bandwidth and linearity of each channel were also experimentally verified using the heterodyne characterization setup (see Supplementary Note \textcolor{magenta}{VI}). Unlike Kerr microcombs \cite{lukashchuk2022dual,riemensberger2020massively,shu2023submilliwatt}, which face degradation in chirp bandwidth due to the Raman effect and higher-order dispersion \cite{karpov2016raman,cherenkov2017dissipative,yi2017single}, our parallel chirp source provides superior chirp uniformity. As shown in Fig.\ref{fig2}f, the chirp bandwidth of the two electro-optic combs (spaced at 21 GHz and 43.6 GHz) exhibits excellent consistency, with channel-dependent chirp bandwidth deviation less than 0.054\% and a maximum deviation of 2.2 MHz. Additionally, channel-dependent nonlinearity remains below 0.2\%, with a maximum RMS (root mean square) below 3 MHz.\\
\indent In summary, our parallel FMCW LiDAR source provides wide spectral tuning range and flexible channel spacing, enabling both broad field coverage and fine resolution in adjustable ROIs. This allows parallel coherent LiDAR to intelligently optimize imaging patterns to meet the varying granularities required in complex scenes. Furthermore, the excellent chirp consistency across channels ensures precise measurement results for each channel, enhancing detection accuracy within the global FOV.

%Fig4
\begin{figure*}[ht]
\centering
\includegraphics[width = 18cm]{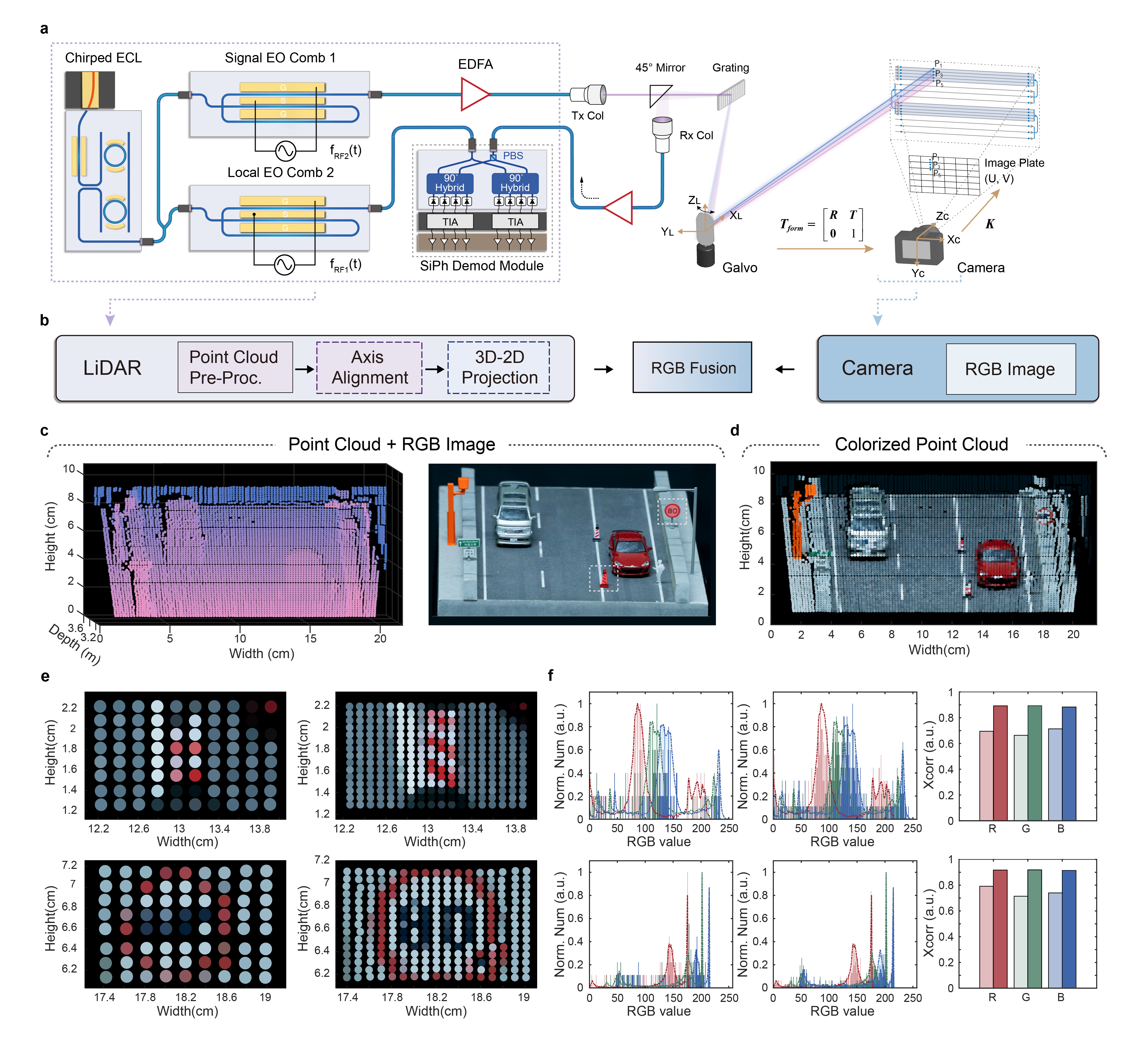}
\caption{\textbf{Adaptive coherent LiDAR with enhanced RGB point cloud visualization. a,} Experimental setup of the RGB-enhanced adaptive coherent LiDAR system, capturing synchronized point cloud data from the LiDAR and an RGB image from a high-resolution camera. PBS: polarization beam splitter. \textbf {b,} Fusion process of LiDAR and camera data. \textbf{c,} Originally processed LiDAR point cloud and RGB image of a simulated road scene with a blackboard background. \textbf{d,} Global colorized point cloud under a standard scan. \textbf{e,} Colorized point cloud of the road sign and barricade under standard scan (left) and locally densified scan (right). \textbf{f,} RGB histogram of the cropped images (envelopes marked by dashed line) and colorized point cloud in \textbf{e} (presented as bars) for the road sign (top) and barricade (bottom), respectively, with the right panel showing the cosine similarity between the image and the point cloud.}
\label{fig4}
\end{figure*}

\vspace{6pt}
\noindent\textbf{Parallel dynamic gaze imaging beyond retina resolution}\\
\noindent To validate the effectiveness of our adaptive LiDAR source, we conduct a proof-of-principle parallel gaze imaging experiment. A linearized triangular frequency modulated signal with a chirp repetition frequency of 100 kHz and a 3.82 GHz chirp bandwidth is used to allow rapid detection capabilities. The signal and local oscillator combs have repetition rates differentiated by 500 MHz to ensure distinct identification of each channel during parallel processing. After optical amplification and collimation, a diffraction transmission grating is used to map the signal comb channels onto pixels in the vertical direction, while a single-axis galvanometer provides the beam scanning in the horizontal direction. The received echoes and local oscillators are coupled into the coherent receiver for multi-heterodyne detection.\\ 
\indent A 3D-printed road scene, positioned approximately 3.2 meters in front of the Galvo mirror, includes one truck, a pedestrian, a barricade, and a box, all variably sized and placed at different spatial depths to simulate real-world conditions. With a sweep of 71 horizontal angles, our LiDAR system constructs point cloud images with resolutions of 54 × 71 pixels for a standard full-FOV scan and 17 × 71 pixels for a locally dense scan centered at any vertical angle, effectively balancing broad FOV coverage and detailed imaging. Limited by the gain spectrum of the optical amplifier, 54 comb lines spaced at 43.5 GHz were used for global detection, managed by only 9 physical channels (see Fig.\ref{fig3}b). The resulting point cloud (as shown in Fig.\ref{fig3}d) clearly distinguishes background, road surface, and targets, with recognizable contours and minimal distortion.\\
\indent We perform dynamic gaze imaging in a target ROI with densified scanning patterns along the spectrum axis, by re-positioning the central wavelength of the frequency chirped ECL and using electro-optic combs with smaller repetition rates. As shown in Fig.\ref{fig3}c, two electro-optic combs with 20.4 GHz channel spacing are utilized to focus on the regions. The focused scan, as shown in the lower inset of Fig.\ref{fig3}d, exhibits a twofold improvement in vertical resolution, revealing more intricate details beneficial for advanced tasks such as point cloud denoising and object classification. The optimal angular resolution in the ROI can achieve 0.012 degrees, which surpasses the requirement of retina imaging (typically 1 arc minute, 0.017 degrees). Such camera-level image capability could provides up to 15 times the resolution of conventional 3D LiDAR sensors. The precision of each point is quantified by the RMSE (root mean square error) of all valid chirp periods. Notably, 90\% of the points exhibit a precision of less than 1.3 cm, with the average precision across the frame being 0.9 cm, as fitted by the Log-Normal distribution model (refer to Supplementary Note \textcolor{magenta}{VII}). With a pixel accumulation time of 10 $\mu$s, our adaptive LiDAR allows for a maximum acquisition rate of 1.7 million pixels per second. Overall, an equivalent 115-line LiDAR, calculated by dividing the FOV by the optimal angular resolution, has been demonstrated using one set of the electro-optic comb with moderate parallelism, which is comparable to current commercial LiDARs.

%Fig5
\begin{figure*}[ht]
\centering
\includegraphics[width = 18cm]{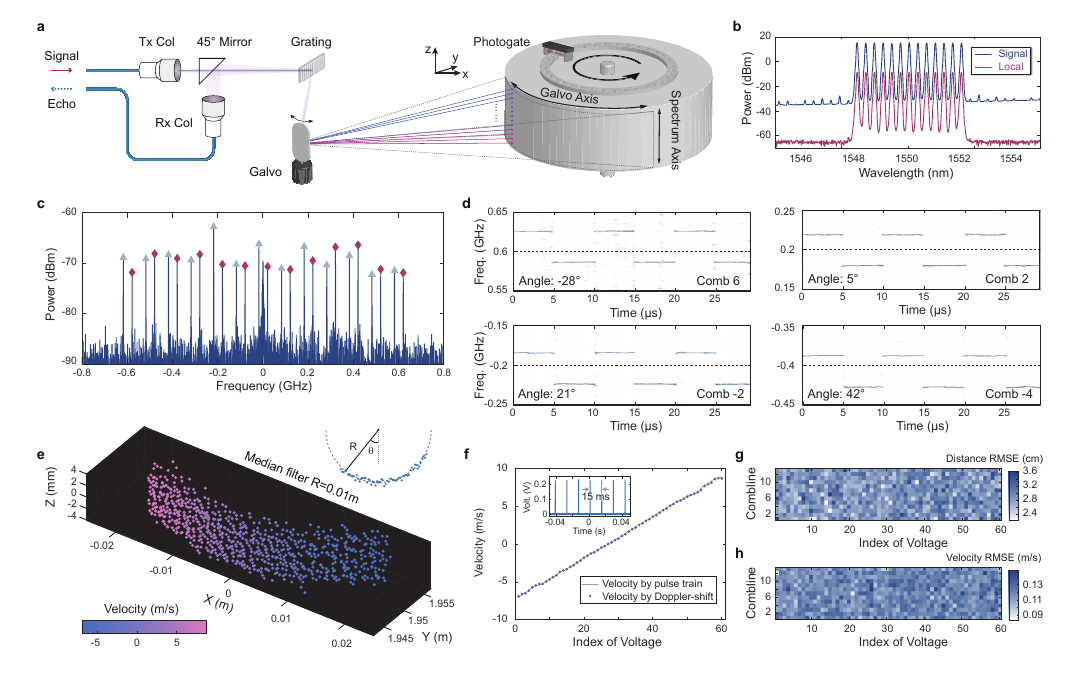}
\caption{\textbf{Parallel 4D imaging experiment.} \textbf{a,} Experiment setup of free space transceiver link and the flywheel target. \textbf{b,} Optical spectra of the emitted signal comb and local oscillators. \textbf{c,} Power spectrum of the single-period complex IQ signal of the coherent receiver, highlighting the peak points.\textbf{d,} Time-frequency map showing various offsets of the mean beat frequency of $\mu\Delta f_{rep}$ due to the Doppler shift. \textbf{e,} Point cloud of the rotating flywheel with the Y-direction speed extracted from the power spectra of the IQ signal. \textbf{f,} Velocities versus various Galvo voltages measured from Doppler shifts, marked by blue points, compared with velocities calculated from the rotating rate extracted from the photogate pulse train (inset) and the angular position of the flywheel extracted from the point cloud. The Y-direction velocities derived from both methods match well. \textbf{g,h} Precision of distance and velocity using single period accumulation time.}
\label{fig5}
\end{figure*}

\vspace{6pt}
\noindent\textbf{Colorized bionic machine vision}\\
\noindent Nowadays, the resolution gap between LiDAR and camera systems is posing significant challenges for achieving a seamless cooperative sensing process to map the precise 3D information of the environment. Here, leveraging the flexible gaze function with ultra-resolution, we further integrate our adaptive coherent LiDAR with a camera to achieve colorized bionic machine vision. In this demo, both LiDAR transmitter and receiver are realized using photonic chip solutions as detailed in Fig.\ref{fig2}. As shown in Fig.\ref{fig4}a, multi-channel probing signals of varying granularity were generated through the collaboration between the hybrid integrated ECL and TFLN chip. To enhance multi-heterodyne detection, a center-punched 45-degree mirror was employed in the bistatic setup to separate the transmitting and echo light paths, effectively eliminating undesired reflections. The SiPh dual-polarization coherent IQ receiver captured both polarized states of scattered light, ensuring stable detection across all channels despite polarization-induced SNR fluctuations.\\ 
\indent To generate colorized point clouds, our adaptive LiDAR was calibrated with the camera to ensure alignment between the captured point cloud and RGB image pixels. The original point cloud was converted into Cartesian coordinates, followed by joint calibration to derive the transformation matrix $T_{form}$ and the camera intrinsic parameters $K$, which allows us to accurately reproject the 3D point cloud onto the 2D image plane (Fig.\ref{fig4}b). Detailed sensor fusion processes are provided in Supplementary Note \textcolor{magenta}{VIII}.\\
\indent Figure \ref{fig4}c shows the RGB image of the simulated road scene and the standard reconstructed point cloud with a resolution of 63 × 101 pixels under a standard full-FOV scan. Our calibrated cooperative imaging leads to a colorized point cloud depicted in Fig.\ref{fig4}d, which clearly captures the background road and two cars with accurate and sufficiently detailed depth and color information. However, low-lying objects such as traffic cones may have a lower detection probability during standard scan due to inadequate point cloud coverage. This can be addressed by locally densifying the point cloud through dynamic gazing. By adjusting the chirp seed wavelength, electro-optic comb line spacing and Galvo mirror, two areas (\ref{fig4}c, white dotted box) where the traffic cone and the road sign are located are assigned denser point clouds than those in the standard scan. From the comparison of the colored point cloud under standard and focused scan (Fig.\ref{fig4}e), we can intuitively see the enhancement in image resolution with finer details, especially the number "80" on a road sign that was not visible in the standard scan. Moreover, Fig.\ref{fig4}g presents the RGB histogram comparisons between cropped images and the colorized point cloud. The cosine similarity (see Supplementary Note \textcolor{magenta}{IX}) between these histograms shows about a 10\% increase, indicating that densified point cloud more closely matches the original image’s color data. The improved alignment between 3D points and image pixels provides enhanced distinction and contextual awareness of key objects, aiding in better cooperative sensing and robust decision-making. Further precision analysis of the point cloud is discussed in Supplementary Note \textcolor{magenta}{X}.

\vspace{6pt}
\noindent\textbf{Parallel 4D imaging}\\
\noindent In a follow-up experiment, parallel 4D imaging is conducted using the two chirped electro-optic comb with comb line spacing of 32.9 GHz and 33 GHz. The system operates at a modulation rate of 100 kHz and a chirp bandwidth of approximately 4 GHz. The emitted signal comb is dispersed by the same transmission grating used in the experiment aforementioned along the vertical direction and swept horizontally across the target by the Galvo mirror, as illustrated in Fig.\ref{fig5}a. A flywheel with a radius $r$ of 2.3 cm rotating at about 67 revolutions per second (RPS) is utilized to simulate a moving object. The complex frequency spectrum derived from sampled data for a single period containing all 13 channels is illustrated in Fig.\ref{fig5}c, with the target peaks during the up- and down-ramps marked in red and blue gray. Detailed time-frequency mappings at different angles and comb lines are depicted in Fig.\ref{fig5}d, where the dashed lines indicate the center frequency of multi-heterodyne detection for each channel. Various Doppler shift frequencies can be observed across different angles. Fig.\ref{fig5}e presents the 4D imaging result of the rotating flywheel, revealing the velocities in the Y direction and spatial positions after applying a median filter to mitigate the influence of the spinning mechanical vibration. The Y-direction velocities of the flywheel can be theoretically calculated from the rotation rate $n$ and the angular position $\theta$ of the flywheel derived from the distance point cloud, using the equation of $v_{Y}=nrsin\theta$, showing strong agreement with the experimental results (inset of Fig.\ref{fig5}e). The rotation rate $n$ can be extracted from the pulse train generated by a photogate on the flywheel, as shown in the inset of Fig.\ref{fig5}f. The precision of the distance and velocity measurements is quantified in Fig.\ref{fig5} g, h, generating an RMSE of 3 cm (root mean square error) for distance and 10 cm / s for the velocity. The results of the velocity measurement of the 90-degree rotating flywheel and using the ECL as the chirp source are shown in the Supplementary Note \textcolor{magenta}{XI}.

\vspace{6pt}
\noindent
\textbf{Discussion} \\
In summary, we have demonstrated an integrated bionic LiDAR that can adaptively gaze at selected ROIs with high resolution, excellent scalability and tunability. These core functionalities are enabled by the efficient and coherent electro-optic interactions in TFLN platform. The generated chirp electro-optic comb source exhibits intrinsic mutual coherence between all comb lines, broadband uniformity, and convenient tunability of the repetition rate and center frequency. This results in massive scanning lines realized with a moderate number of physical channels, which is of practical significance in reducing system size and cost. Furthermore, the broadband tunability of the ECL wavelength enables solid-state positioning of the ROI anywhere within the FOV and supports continuous scanning for small obstacles. Adaptive parallel detection is facilitated through multi-heterodyne detection of two sets of electro-optic combs driven by the same chirped source. As the relative frequency difference between comb lines remains constant, changes in wavelength during spectral scanning do not impede the parallel processing of all channels.\\
\indent Despite that four independent chips are utilized in the conceptual demo, our LiDAR system holds significant potential for full integration on a TFLN platform: integrated Pockels laser \cite{li2022integrated,franken2024high} can supply a rapidly chirped and spectrally scanned source; the gain and saturated output power of rare-earth-doped TFLN amplifiers \cite{liu2022photonic,zhou2021chip,wang2024erbium,zhang2023highly,cai2021erbium} are approaching the performance of commercial fiber amplifiers; Additionally, non-mechanical beam steering is poised to be achieved through optical phased arrays \cite{sun2013large,liu2022silicon,chen2024sin,xu2019aliasing} or arrays of acousto-optic devices\cite{li2023frequency}. Although the pure LN platform lacks native light sources and photodetectors, the heterogeneous integration of TFLN with other materials such as silicon \cite{wei2023ultra,he2019high,xiang2021high}, silicon nitride \cite{churaev2023heterogeneously,snigirev2023ultrafast} or III-V compounds \cite{zhang2022heterogeneously,guo2022high,zhang2023heterogeneous} can provide high-quality active devices.\\
\indent With the ongoing efforts in multi-function and multi-material integration, a fully integrated artificial “macular” is foreseeable. Assuming full utilization of the 100 nm spectral range (1486-1590 nm) with 20.4 GHz channel spacing, a 637-line coherent LiDAR is possible, far surpassing current long-range ToF and FMCW systems. The ROI for scanning can also be expanded using ultra-low $V_{\pi}$ modulators \cite{liu2022broadband,li2023compact,he2024chip}, allowing the application of a higher modulation index to broaden the comb spectrum, while maintaining reasonable RF power levels. Furthermore, the 4D perception enabled by coherent detection provides essential velocity and acceleration data for Simultaneous Localization and Mapping (SLAM), particularly in GPS-denied environments like tunnels.\\
\indent The demonstrated capabilities of reconfigurable imaging patterns could also inspire new applications in robotics. For instance, a heterogeneous eye system akin to that of a jumping spider could be constructed for bionic robotics using our adaptive LiDARs, with each unit tailored with different channel spacing to meet specific imaging resolution requirements in various directions. Moreover, combining several compact LiDAR modules could lead to a beyond-bionic machine vision system that outperforms traditional monocular and compound eye configurations. This setup would divide the global FOV into multiple regions with high acquisition rates and pixel density, while enabling flexible, dynamic gaze scanning across the entire FOV. Additionally, the idea of simultaneously utilizing frequency down-conversion and scalable repetition rate of electro-optic comb can be migrated to fast and fine spectral analysis \cite{shams2022thin,qing2024vector,xu2019broadband,wang2018fast}, which can also lead to impact on adaptable optical communication \cite{lukens2019all}, optical coherence tomography \cite{siddiqui2018high}, compressive sensing and imaging\cite{giorgetta2023free}, and high-resolution optical metrology \cite{udem2002optical,weimann2017silicon,weimann2018fast,tran2024near}.
 
 \vspace{6pt}
\noindent \textbf{Methods}\\
\begin{footnotesize}
\noindent \textbf{Chirp Seed Generation.} 
For frequency-chirped ECL, any section (including the RSOA, the MRRs, and the phase shifter) that changes the laser cavity round-trip phase can be modulated to generate the frequency chirp. To generate a sufficiently large chirp bandwidth, we synchronously drive the MRRs to synchronously tune the Vernier filter spectrum, while the phase shifter is applied for phase compensation. The resonance wavelengths of $MRR_{1}$ and $MRR_{2}$ remain aligned during the frequency chirp. Due to the intracavity mode competition mechanism, the lasing mode should satisfy the resonating condition of the Vernier MRR. Therefore, the lasing longitudinal mode also shifts with the Vernier filter shift. The detuning of them could be compensated via the phase shifter. In resonance, the additional phase introduced to the laser cavity by $MRR_{i}$ is $\varphi_{c,i}=\frac{L_{e,i}}{L_{i}}\varphi_{i}$. The laser wavelength shift $\delta\lambda_{c}$ is related to the ring resonator phase change $\delta\varphi_{i}$ by

\begin{equation}
\begin{aligned}
\delta\lambda_{c} = \sum_{i} \frac{FSR_{laser}}{2\pi} \frac{L_{e,i}}{L_{i}} \delta\varphi_{i}
\end{aligned}
\end{equation}

\noindent According to the enhancement factor $L_{e,i}⁄L_{i}$, the MRR induces a wavelength shift greater than the phase shifter by applying the same phase change. The misalignment between the laser longitudinal mode and the filter passband could be compensated via the phase shifter, leading to a larger chirp bandwidth.

\vspace{3pt}
\noindent\textbf{TFLN Device Fabrication.} 
Our TFLN devices are fabricated from a commercially available x-cut LNOI wafer (4-inch, NANOLN), which includes a 500 nm LN thin film on the top, a 2 $\mu$m buried silicon dioxide layer, and a 500 $\mu$m silicon substrate. The ultraviolet (UV) and $Ar^{+}$-based reactive ion etching (RIE) systems are used to define LN optical waveguides and other structures. The entire device is then cladded with silicon dioxide by plasma-enhanced chemical vapor deposition (PECVD), while metal electrodes are fabricated using a sequence of photolithography, thermal evaporation, and lift-off processes \cite{zhang2023power}. In our design, capacitive loaded traveling wave electrodes are applied for an efficient match between microwave and optical velocity throughout the modulation region \cite{zhang2023power}. Finally, the devices are cleaved and the facets are carefully polished for end-fire fiber-to-chip coupling.     

\vspace{3pt}
\noindent\textbf{Experimental Details.} 
For parallel imaging demo, the full FOV is scanned utilizing six sets of electro-optic combs with central wavelengths spaced at approximately 3 nm. Due to limitations in experimental equipment, we used two RF amplifiers (TLPA18G50G-35-27) with a saturation power of 27 dBm to generate electro-optic combs. Each set of electro-optic combs is flattened and amplified in an erbium-doped fiber amplifier (EDFA) before emitting to the free space, and the average power of one single channel is below 10 dBm, in compliance with eye safety regulations. For Experiment of "Parallel dynamic gaze imaging beyond retina resolution" (Exp. I), a pair of 10mm achromatic lenses are used for beam focusing to minimize spatial transmission loss, while in experiment of "Colorized bionic machine vision" (Exp. II) and of "Parallel 4D imaging" (Exp. III), two zoom fiber collimators are utilized to perform the same function. Afterwards, the multichannel chirped signals are vertically dispersed by a 966 lines/mm transmission grating before being guided to the Galvo (Thorlab QS45Y-AG) for horizontal scanning. 

Due to the utilization of a monostatic scheme, a portion of the light returned from the target returns along part of the original path. In Exp. I, they are collected by the transmitting fiber collimator, and directed to the receiver via the circulator, while in Exp. II\&III, they are guided to an independent optical path using a center-punched 45° mirror. For multi-heterodyne detection, the reflected light of the emitted signal comb is amplified and then coupled into a silicon photonic IQ coherent receiver with the reference comb. Passing through the on-chip polarization beam splitter (PBS) and phase-tuning devices, the signal comb and the reference comb are then mixed in 90 ° optical hybrids and converted to photocurrent by photodetectors. After being amplified by the TIAs, the differential I and Q beat signals of both polarizations are acquired by the broadband Baluns (Marki Microwave BAL0026). The oscilloscope (Keysight MXR404A) can provide a 4 GHz operating bandwidth on four simultaneous channels, representing a low dependence on complex high-speed devices. For signal processing, the fast Fourier transform of complex signal I+iQ reveals the two-sided spectrum that contains information on all parallel comb lines, as shown in Fig. \ref{fig5}c. The collection times for each horizontal angle are set to 625 $\mu$s, 5 ms, and 1 ms in Exp.I, II and III, in order to acquire enough periods for precision analysis.  

\vspace{3pt}
\noindent\textbf{LiDAR-Camera Calibration.} 
To fuse the LiDAR point cloud and camera-colored pixels, we need to estimate the transformation matrix to that gives the relative rotation and translation between the two sensors. We used a custom calibration board with an 8 × 5 checkerboard pattern. Each square of the checkerboard measures precisely 13.35 mm. First, we position the calibration board in the FOV of both the LiDAR and the camera and capture a set of synchronized point-cloud data frames from the LiDAR and images from the camera. A sony camera (6000 × 4000 pixels) was used for the RGB image collection. Then, utilizing computer vision algorithms, we detect the checkerboard corners in the camera images and separate the point cloud corresponding to the calibration board. Therefore, we can use the detected 2D corners in the camera images and the corresponding 3D points from the LiDAR data to compute the transformation between the camera and the LiDAR coordinate systems. We adopted an algorithm from MATLAB LiDAR Toolbox that minimizes the reprojection error \cite{zhou2018automatic} to extract the transformation matrix. Finally, we project the overall point-cloud data onto the image and match the point cloud with image pixels, and the colored point-cloud data are acquired.

\vspace{3pt}
\noindent\textbf{Data availability}\\
%The authors declare that the data supporting the findings of this study are available within the article and its Supplementary Information. All raw data are available from the corresponding author upon reasonable request.
The data that supports the plots within this paper and other findings of this study are available from the corresponding authors upon reasonable request. 

\vspace{3pt}
\noindent\textbf{Code availability}\\
The codes that support the findings of this study are available from the corresponding authors upon reasonable request.
\end{footnotesize}
\vspace{20pt}

%_______REFERENCE____________%
%\bibliography{Lidar.bib}\\
% \bibliographystyle{naturemag}
% \bibliographystyle{naturesaa}

%%_______REFERENCE_______%%

%--------------------------------------------------------------------------
\vspace{12pt}
\begin{footnotesize}

\vspace{6pt}
\noindent \textbf{Acknowledgment}

\noindent 
This work was supported by National Key R\&D Program of China (2021YFB2800400), National Natural Science Foundation of China under Grant (62001010, 62322501, 12204021), China National Postdoctoral Program for Innovative Talents (BX20240014), Research Grants Council, University Grants Committee (CityU 11212721, C1002-22Y), Croucher Foundation (9509005) and High-performance Computing Platform of Peking University. We thank Chen Wang for fruitful discussions. We thank C. F. Yeung, S. Y. Lao, C. W. Lai and L. Ho at the Nanosystem Fabrication Facility at the Hong Kong University of Science and Technology for technical support with the stepper lithography and plasma-enhanced chemical vapour deposition process. We thank W. H. Wong and K. Shum at CityU for their help in device fabrication and measurement. The CityU nano-fabrication facility was used.\\

\noindent \textbf{Author contributions}\\
\noindent The experiments were conceived by R.C., Y.W. and C.L. The devices were designed by K.Z., Y.C., and C.L. The TFLN devices were fabricated by Z.C., H.F. LiDAR-camera calibration was conducted by W.L. Electro-optic packaging was conducted by Z.G. and Y.Z. Other characterizations were conducted by R.C., Y.W. and C.L., with the assistance from Y.C., W.L., B.S., Z.T. and Y.W. The results were analyzed by R.C., Y.W., and K.Z. All authors participated in writing the manuscript. The project was coordinated by K.Z. and W.X. under the supervision of H.S., L.Z., C.W. and X.W.\\ 

\noindent
\textbf{Additional information}\\
\noindent Supplementary information is available in the online version of the paper. Reprints and permissions information is available online. Correspondence and requests for materials should be addressed to C.W. and X.W.

\vspace{6pt}
\noindent \textbf{Competing interests} 

\noindent K.Z., Z.C., H.F. and W.C. are involved in developing lithium niobate technologies at Kokoxili Photonics Limited.
\end{footnotesize}
%--------------------------------------------------------------------------

\end{document}